# Velocity distribution of high-energy particles

Jian-Miin Liu*
Department of Physics, Nanjing University
Nanjing, The People's Republic of China
*On leave. Present mailing address: P.O.Box 1486, Kingston, RI 02881, USA.

ABSTRACT
High energy infers high velocity and high velocity is a concept of special relativity. The Maxwellian velocity distribution is corrected to be consistent with special relativity. The corrected velocity distribution reduces to the Maxwellian distribution for small velocities and vanishes at the velocity of light, and contains a relatively depleted high-energy tail.

**Introduction.** The deviation of velocity distribution of high-energy particles from the Maxwellian distribution has been observed for many years [1-4]. For this, $\kappa$ (kappa) velocity distribution

$$\frac{N}{\pi^{3/2}} \frac{1}{\theta^3} \frac{\Gamma(\kappa+1)}{\kappa^{3/2}\Gamma(\kappa-1/2)} (1+\frac{y^2}{\kappa\theta^2})^{-(\kappa+1)}$$

with $\theta = [(2\kappa-3)/\kappa]^{1/2}(K_B T/m)^{1/2}$ and its alterations were suggested [2-4], where $\Gamma$ is the gamma function, $y^r$ is the velocity magnitude, $y=(y^r y^r)^{1/2}$, r=1,2,3, and kappa is a parameter to be determined in comparison with experimental data. Here, different values of kappa correspond to different kinds of velocity distribution. When and only when kappa goes to infinity, the kappa distribution becomes the Maxwellian. Kappa distribution and its alterations are rather phenomenological: The kappa values in fitting experimental data vary from event to event [2].

High energy infers high velocity and high velocity is a concept of special relativity. Therefore, for high-energy particles, we need to look for an equilibrium velocity distribution which is consistent with Special Relativity (SR-consistent, hereafter). Kappa distribution and its alterations are not SR-consistent as they are non-vanishing for velocities greater than the velocity of light.

In this letter, we show that although we have not had an acceptable Lorentz-invariant statistical mechanics due to the sharp conflict between the concepts in statistical mechanics based on pre-relativistic mechanics and those in special relativity [5], from which we can deduce a SR-consistent equilibrium velocity distribution, we still can have such a velocity distribution through analyzing velocity space.

**Velocity space.** Velocity space is a space in which point pairs represent relative velocities. The three-dimensional velocity space defined by

$$dY^2 = H_{rs}(y)dy^r dy^s, \quad r,s=1,2,3, \tag{1a}$$

$$H_{rs}(y) = c^2\delta^{rs}/(c^2-y^2) + c^2 y^r y^s/(c^2-y^2)^2, \quad \text{real } y^r \text{ and } y<c, \tag{1b}$$

in the usual velocity-coordinates $\{y^r\}$, r=1,2,3, where $y^r$ is the well-defined Newtonian velocity, $y=(y^r y^r)^{1/2}$ and c is the velocity of light, has been studied for a long time [6-8]. This velocity space is characterized by a finite boundary at c and the Einstein velocity addition law. Mathematically, this velocity space can be represented in terms of the so-called primed velocity-coordinates $\{y'^r\}$, r=1,2,3, which are connected with the usual velocity-coordinates by

$$dy'^r = A^r_s(y)dy^s, \quad r,s=1,2,3, \tag{2a}$$

$$A^r_s(y) = \gamma \delta^{rs} + \gamma(\gamma-1) y^r y^s / y^2, \tag{2b}$$

where $\gamma = 1/(1-y^2/c^2)^{1/2}$. The represented velocity space has the Euclidean structure,

$$dY^2 = \delta_{rs} dy'^r dy'^s, \quad r,s=1,2,3, \tag{3}$$

in the primed velocity-coordinates because

$$\delta_{rs} A^r_p(y) A^s_q(y) = H_{pq}(y), \quad r,s,p,q=1,2,3.$$

With standard calculation techniques in Riemann geometry, we can find

$$H^{rs}(y) = (c^2-y^2)\delta^{rs}/c^2 - (c^2-y^2) y^r y^s / c^4,$$

$$\Gamma^i_{jk} = \begin{cases} 2y^i/(c^2-y^2), & \text{if } i=j=k; \\ y^k/(c^2-y^2), & \text{if } i=j \neq k; \\ y^j/(c^2-y^2), & \text{if } i=k \neq j; \\ 0, & \text{otherwise,} \end{cases}$$

where $H^{rs}(y)$ is the contravariant metric tensor and $\Gamma^i_{jk}$ is the Christoffel symbols. The equation of geodesic line is therefore

$$\ddot{y}^r + [2/(c^2-y^2)] \dot{y}^r (y^s \dot{y}^s) = 0, \quad r,s=1,2,3, \tag{4}$$

where dot refers to the derivative with respect to velocity-length. Introducing new variables

$$w^r = \dot{y}^r/(c^2-y^2), \quad r=1,2,3, \tag{5}$$

we are able to rewrite Eq.(4) as

$$\dot{w}^r = 0, \quad r=1,2,3. \tag{6}$$

It is seen that

$$w^r = \text{constant}, \quad r=1,2,3, \tag{7}$$

are a solution to Eqs.(6). Due to Eqs.(5) and (6), we have

$$w^s y^r - w^r y^s = \text{constant}, \quad r,s=1,2,3. \tag{8}$$

Eqs.(7) and (8) specify three linear relations between any two of $y^1$, $y^2$ and $y^3$. These linear relations give the following shape to the equation of geodesic line between two points, $y_1^r$ and $y_2^r$, $r=1,2,3$,

$$y^r = y_1^r + \alpha(y_2^r - y_1^r), \quad 0 \leq \alpha \leq 1, \quad r=1,2,3. \tag{9}$$

Using Eqs.(9), at some length, we can find the velocity-length between two points $y_1^r$ and $y_2^r$,

$$Y(y_1^r, y_2^r) = \frac{c}{2} \ln \frac{b+a}{b-a}, \tag{10a}$$

$$b = c^2 - y_1^r y_2^r, \quad r=1,2,3, \tag{10b}$$

$$a = \{(c^2 - y_1^i y_1^i)(y_2^j - y_1^j)(y_2^j - y_1^j) + [y_1^k(y_2^k - y_1^k)]^2\}^{1/2}, \quad i,j,k=1,2,3. \tag{10c}$$

In the case of $y_1^r = 0$ and $y_2^r = y^r$, the velocity-length is

$$Y(0,y^r) = \frac{c}{2} \ln \frac{c+y}{c-y} \quad \text{or} \quad Y^2(0,y^r) = [\frac{c}{2y} \ln \frac{c+y}{c-y}]^2 \delta_{rs} y^r y^s, \quad r,s=1,2,3. \tag{11}$$

On the other hand, we know from Eqs.(3) that square of velocity-length between two points $y_1'^r$ and $y_2'^r$ is

$$Y^2(y_1'^r, y_2'^r) = \delta_{rs}(y_2'^r - y_1'^r)(y_2'^s - y_1'^s), \quad r,s=1,2,3, \tag{12}$$

and

$$Y^2(0, y'^r) = \delta_{rs} y'^r y'^s, \quad r,s=1,2,3, \tag{13}$$

if $y_1'^r = 0$ and $y_2'^r = y'^r$. Eqs.(11) and (13) imply

$$y'^r = [\frac{c}{2y} \ln \frac{c+y}{c-y}] y^r, \quad r=1,2,3, \tag{14}$$

$$y' = \frac{c}{2} \ln \frac{c+y}{c-y}, \tag{15}$$

when $(y'^1, y'^2, y'^3)$ and $(y^1, y^2, y^3)$ represent the same point in the velocity space, where $y' = (y'^r y'^r)^{1/2}$, $r=1,2,3$.

We call $y'^r$, $r=1,2,3$, the primed velocity [9]. Its definition from the measurement point of view is given in Ref.[10]. The Galilean addition law of primed velocities links up with the Einstein addition law of corresponding Newtonian velocities [9,10].

**Equilibrium velocity distribution.** The Euclidean structure of the velocity space in the primed velocity-coordinates, Eq.(3), convinces us that the Maxwellian velocity and velocity rate distribution formulas are valid in the primed velocity-coordinates, namely

$$P(y'^1,y'^2,y'^3)dy'^1dy'^2dy'^3 = N\left(\frac{m}{2\pi K_B T}\right)^{3/2} \exp\left[-\frac{m}{2K_B T}(y')^2\right]dy'^1dy'^2dy'^3 \tag{16}$$

and

$$P(y')dy' = 4\pi N\left(\frac{m}{2\pi K_B T}\right)^{3/2}(y')^2 \exp\left[-\frac{m}{2K_B T}(y')^2\right]dy', \tag{17}$$

where N is the number of particles, m their rest mass, T the temperature, and $K_B$ the Boltzmann constant.

We can employ Eqs.(2a-2b) and (15) to represent these two formulas in the usual velocity-coordinates $\{y^r\}$, r=1,2,3. Using Eq.(15) and

$$dy'^1dy'^2dy'^3 = \gamma^4 dy^1dy^2dy^3$$

which is inferred from Eqs.(2a-2b), we have from Eq.(16),

$$P(y^1,y^2,y^3)dy^1dy^2dy^3 = N\frac{(m/2\pi K_B T)^{3/2}}{(1-y^2/c^2)^2}\exp\left[-\frac{mc^2}{8K_B T}\left(\ln\frac{c+y}{c-y}\right)^2\right]dy^1dy^2dy^3. \tag{18}$$

Using Eq.(15) and

$$dy' = \gamma^2 dy$$

which comes from differentiating Eq.(15), we have from Eq.(17),

$$P(y)dy = \pi c^2 N\frac{(m/2\pi K_B T)^{3/2}}{(1-y^2/c^2)}\left(\ln\frac{c+y}{c-y}\right)^2 \exp\left[-\frac{mc^2}{8K_B T}\left(\ln\frac{c+y}{c-y}\right)^2\right]dy. \tag{19}$$

**Corrections to the Maxwellian velocity and velocity rate distributions.** For small velocities, velocity distribution function $P(y^1,y^2,y^3)$ and velocity rate distribution function $P(y)$ respectively reduce to

$$N\left(\frac{m}{2\pi K_B T}\right)^{3/2}\exp\left[-\frac{m}{2K_B T}(y^2)\right] \tag{20}$$

and

$$4\pi N\left(\frac{m}{2\pi K_B T}\right)^{3/2}(y^2)\exp\left[-\frac{m}{2K_B T}(y^2)\right], \tag{21}$$

which are just the Maxwellian velocity and velocity rate distribution functions. $P(y^1,y^2,y^3)$ and $P(y)$ become nothing when y is greater than c. They vanish as y approaches c. To prove, we write

$$\lim_{y\to c} P(y^1,y^2,y^3) = \lim_{z\to+\infty}\frac{c^2 N}{4}\left(\frac{m}{2\pi K_B T}\right)^{3/2}z^2\exp\{-A[\ell n(2cz)]^2\},$$

$$\lim_{y\to c} P(y) = \lim_{z\to+\infty}\frac{\pi c^3 N}{2}\left(\frac{m}{2\pi K_B T}\right)^{3/2}z[\ell n(2cz)]^2\exp\{-A[\ell n(2cz)]^2\},$$

where $z=1/(c-y)$ and $A=\frac{mc^2}{8K_B T}$. Since $\ell n(2cz)$ is smaller than $2cz$ for large z, both $\lim_{y\to c} P(y^1,y^2,y^3)$ and $\lim_{y\to c} P(y)$ are smaller than $\lim_{z\to+\infty}(\text{constant})z^3\exp\{-A[\ell n(2cz)]^2\}$. The last limit equals zero. Eqs.(18) and (19) are the corrections to the Maxwellian velocity and velocity rate distributions.

The positive and monotonically decreasing high-energy tail of velocity distribution function $P(y^1,y^2,y^3)$ goes to zero as y approaches c, while the positive and monotonically decreasing high-energy tail of the Maxwellian velocity distribution function in Eq.(20) goes to zero as y approaches infinity. That indicates a depleted high-energy tail of the corrected velocity distribution with respect to the Maxwellian velocity distribution. The same situation exists between velocity rate distribution function P(y) and the Maxwellian velocity rate distribution function.

**Conclusion.** We have analyzed the velocity space. As a result, we obtain the SR-consistent equilibrium velocity and velocity rate distributions. They reduce to the Maxwellian velocity and velocity rate distributions for small velocities and vanish at the velocity of light. They each contains a relatively depleted high-energy tail.

The corrected velocity and velocity rate distributions differ from the Maxwellian velocity and velocity rate distributions substantially in the part of high velocity or high energy. Besides the deviation from the Maxwellian distribution showing in velocity distributions of high-energy particles in planetary magnetospheres and solar wind, we can expect their non-negligible effects in the phenomena concerning with those statistical calculations that are mainly relevant to the high-energy part of velocity and velocity rate distributions or those statistical calculations where most particles crowd in the high-energy part. Some applications of the corrected velocity and velocity rate distributions will be reported elsewhere [11].

ACKNOWLEDGMENT
   The author greatly appreciates the teachings of Prof. Wo-Te Shen. The author thanks Prof. Gerhard Muller and Dr. P. Rucker for useful suggestions.